%% file: paper.tex
\documentclass[secnumarabic,aps,pra,amsfonts,twocolumn,%
showkeys
]{revtex4-1}

\usepackage{bm,amsfonts,amsmath,amsthm}
\usepackage{placeins}
\usepackage{longtable}
\usepackage{graphicx}
\newcommand{\gl}[1]{(\ref{#1})}

\begin{document}

\title{{\em k}\! -space magnetism as the origin of superconductivity}
\author{Ekkehard Kr\"uger} 
\affiliation{Institut f\"ur Materialwissenschaft, Materialphysik,
  Universit\"at Stuttgart, D-70569 Stuttgart, Germany}
%
\date{\today}

\begin{abstract}
The nonadiabatic Heisenberg model presents a nonadiabatic
  mechanism generating Cooper pairs in narrow, roughly half-filled
  ``superconducting bands'' of special symmetry. Here we show that this
  mechanism may be understood as the outcome of a special spin
  structure in the reciprocal space, hereinafter referred to as
  ``$k$-space magnetism''. The presented picture permits a vivid
  depiction of this new mechanism highlighting the height similarity
  as well as the essential difference between the new nonadiabatic and
  the familiar Bardeen-Cooper-Schrieffer mechanism.
\end{abstract}

\maketitle

\section{Introduction}
\label{sec:introduction}
The nonadiabatic Heisenberg model (NHM)~\cite{enhm} is an extension of
the Heisenberg model~\cite{hei} going beyond the adiabatic
approximation. It is based on three postulates related to the
atomic-like motion~\cite{hei,mott,hubbard} of the electrons in narrow,
roughly half-filled energy bands. An atomic-like motion is
characterized by electrons occupying localized states which for their
part move as Bloch waves through the crystal. The NHM does not
represent the localized states by (hybrid) atomic functions but
solely by symmetry-adapted and optimally-localized Wannier functions
forming an {\em exact} unitary transformation of the Bloch functions
of a narrow, roughly half-filled energy band.

The energy bands in the band structures of the metals are degenerate
at several points and lines (of symmetry) of the Brillouin
zone. Hence, it is generally not possible to find narrow, roughly
half-filled {\em closed} energy bands in the band structures of the
metals as they are required for the construction of optimally
localized Wannier functions. However, in the band structures of those
metals that experimentally prove to be superconductors, the
construction of such Wannier functions becomes possible if we allow
the Wannier functions to be spin dependent~\cite{theoriewf}. This
observation leads to the definition of ``superconducting
bands''~\cite{theoriewf}: The Bloch functions of a superconducting
band can be unitarily transformed into optimally localized
spin-dependent Wannier functions that are symmetry-adapted to the full
space group of the metal.


Within the NHM, the atomic-like motion in a superconducting band
produces Cooper pairs below a transition temperature~\cite{josn}. The
aim of the paper is to show that this nonadiabatic mechanism can be
understood as the outcome of a special spin structure in the
reciprocal space referred to as ``$\bm k$-space magnetism''. In
Section~\ref{sec:ksm} we shall declare what we mean by $\bm k$-space
magnetism. In Section~\ref{sec:cooperpairs} we will show
that $\bm k$-space magnetism leads directly to the formation of Cooper
pairs at low temperatures, and in Section~\ref{sec:nhm}, finally, we
will show that the NHM provides an interaction producing $\bm k$-space
magnetism in narrow, roughly half-filled superconducting bands when we
leave the adiabatic approximation.

\section{{\em k}-space magnetism}
\label{sec:ksm}
Within the NHM, strongly correlated electrons in a narrow, roughly
half-filled superconducting band produce a special spin structure at
the Fermi level that we call ``$\bm k$-space magnetism'': the electron
spins of the Bloch electrons are no longer parallel or anti-parallel
to a fixed symmetry axis (usually the $z$ axis), but are parallel or
anti-parallel to an axis $z_{\bm k}$ determined by the $\bm k$ vector
of the electron, as is visualized in Figure~\ref{fig:zeichnung}. The
direction of $z_{\bm k}$ changes continuously in the $\bm k$ space and
is not independent of $\bm k$ in a narrow, roughly half-filled
superconducting band. The spin $s = \pm\frac{1}{2}$ of the Bloch
electron at wave vector $\bm k$ still may lie parallel (for $s =
+\frac{1}{2}$) or anti-parallel (for $s = -\frac{1}{2}$) to the
predefined $z_{\bm k}$ axis. Thus, $\bm k$-space magnetism does not
create a magnetic field and is invariant under time inversion.


\input{Figur}

Two questions emerge at this point: first, why $\bm k$-space magnetism
produces Cooper pairs, and, secondly, which interaction produces $\bm
k$-space magnetism. These questions shall be answered in the
Sections~\ref{sec:cooperpairs} and~\ref{sec:nhm}, respectively.

\section{{\em k}-space magnetism
 producing Cooper pairs}
\label{sec:cooperpairs}
Consider an electron system ${\cal E}^{km}$ exhibiting $\bm k$-space
magnetism at the Fermi level.  The interaction producing the $\bm
k$-space magnetism shall be defined in the
Section~\ref{sec:nhm}, here we assume it to exist.

At any scattering process in the electron system ${\cal E}^{km}$ the
total electron spin of the scattered electrons is not conserved since
the spin direction is $\bm k$ dependent. Hence, the electrons must
interchange spin angular momenta with the lattice of the atomic
cores. As a consequence (Section 3.1 of Ref.~\cite{josn}), at any
electronic scattering process two crystal-spin-1 bosons are excited or
absorbed.

At zero (or very low) temperature the crystal-spin-1 bosons will be
only virtually excited. That means that each boson pair is reabsorbed
instantaneously after its generation. Hence, whenever a boson pair is
excited during a certain scattering process
\begin{equation}
  \label{eq:7}
\bm k_1, \bm k_2 \rightarrow\bm k_1', \bm k_2'  
\end{equation}
of the two electrons $\bm k_1$ and $\bm k_2$, this boson pair is
reabsorbed instantaneously during a second scattering process
\begin{equation}
  \label{eq:8}
\bm k_3, \bm k_4 \rightarrow\bm k_3', \bm k_4'  
\end{equation} 
of two other electrons $\bm k_3$ and $\bm k_4$.
Consequently, the resulting total scattering process 
\begin{equation}
  \label{eq:9}
\bm k_1, \bm k_2,\bm k_3, \bm k_4 \rightarrow\bm k_1', \bm k_2',\bm
k_3', \bm k_4' 
\end{equation}
must conserve the total electron spin. Only in this case, the boson
pair created during the first process \gl{eq:7} is completely
reabsorbed during the second process \gl{eq:8}. However, also at
the scattering processes \gl{eq:9} of four electrons, the total spin is
generally not conserved since the spin direction still is $\bm k$
dependent.

The only scattering processes within ${\cal E}^{km}$ conserving the
total electron spin are scattering processes between Cooper pairs:
since the system is invariant under time-inversion, the spins of the
Bloch states labeled by $\bm k$ and by $-\bm k$ lie exactly
opposite. When both states are occupied at the same time, they form a
Cooper pair with exactly zero total spin. Hence, any scattering
process between Cooper pairs
\begin{equation}
  \label{eq:10}
  \bm k_1, -\bm k_1;\ \bm k_2, -\bm k_2 \rightarrow\bm k_1', -\bm
  k_1';\ \bm k_2', -\bm
  k_2' 
\end{equation}
conserves the total spin angular momentum within ${\cal E}^{km}$, see
the detailed group-theoretical discussion in Section 3.2 of
Ref.~\cite{josn}.  This scattering process~\gl{eq:10} comprises the
two processes
\begin{equation}
\bm k_1, \bm k_2 \rightarrow\bm k_1', \bm k_2'  
\end{equation}
destroying a Cooper pair and creating a boson pair, and the subsequent 
process
\begin{equation}
  \label{eq:2}
-\bm k_1, -\bm k_2 \rightarrow -\bm k_1', -\bm k_2'
\end{equation}
recomposing the Cooper pair and reabsorbing the boson pair. This only
possible combined scattering process within ${\cal E}^{km}$ represents
the well-known Bardeen-Cooper-Schrieffer (BCS) mechanism~\cite{bcs} in
${\cal E}^{km}$, see Section 3.2 of Ref.~\cite{josn}. However, the
mechanism in ${\cal E}^{km}$ differs from the BCS mechanism because it
is effective {\em solely} between Cooper pairs. It {\em necessarily}
produces Cooper pairs possessing only one half of the degrees of
freedom of free electrons. This necessary reduction of the degrees of
freedom may be compared with the effect of constraining forces in
classical systems. Thus, we speak of quantum mechanical constraining
forces stabilizing the Cooper pairs in ${\cal E}^{km}$~\cite{josn},
or, more illustratively, by ``spring mounted Cooper
pairs''~\cite{josi}.

\section{Strongly correlated electrons producing {\em k}-space magnetism}
\label{sec:nhm}
In the framework of the NHM, the electrons of a narrow, roughly
half-filled superconducting band lower their total Coulomb energy
by producing $\bm k$-space magnetism. This far-reaching assertion
follows from the three postulates of the NHM~\cite{enhm} and from the
special properties of the spin-dependent Wannier functions
representing the atomic-like states in a superconducting band. In
Section~\ref{sec:sdwf} we first shall repeat the definition of
spin-dependent Wannier functions in the special case of a metal with
one atom in the unit cell (the general definition is given in
Ref.~\cite{theoriewf}), and in Section~\ref{sec:interactionksm} we
shall show that the postulates of the NHM define an interaction
producing $\bm k$-space magnetism.

\subsection{Spin-dependent Wannier functions}
\label{sec:sdwf}
For the sake of simplicity, we consider a metal with only one atom in
the unit cell. In this case, superconducting bands are single
bands~\cite{theoriewf}. Furthermore, we assume that this metal
possesses a narrow, half-filled superconducting band in its band
structure. By definition we can unitarily transform the Bloch
functions of this band into optimally localized and symmetry-adapted
spin-dependent Wannier functions~\cite{theoriewf}.  We do this by
replacing the Bloch functions $\varphi_{\bm{k}}(\bm{r})$ of the
superconducting band by Bloch spinors
\begin{equation}
  \label{eq:1}
  \varphi_{\bm{k},m}(\bm{r},t) = \sum_{s = -\frac{1}{2}}^{+\frac{1}{2}}
  f_{ms}(\bm{k})u_s(t)\varphi_{\bm{k}}(\bm{r})
\end{equation}
with $\bm k$ dependent spin directions. The functions $u_{s}(t)$
denote Pauli's spin functions:
\begin{equation}
 \label{eq:3}
 u_{s}(t) = \delta_{st},
\end{equation}
where $s = \pm\frac{1}{2}$ and $t = \pm\frac{1}{2}$ are the spin
quantum number and the spin coordinate, respectively. (To simplify,
we ignore that in some points of symmetry the Bloch spinors may not be
written in the form~\gl{eq:1}~\cite{theoriewf}.)  The coefficients
$f_{ms}(\bm{k})$ in Equation~\gl{eq:1} form a $\bm k$ dependent
two-dimensional matrix
\begin{equation}
 \label{eq:94}
 {\bf f}(\bm{k}) = [f_{ms}(\bm{k})] 
\end{equation}
which is unitary,
\begin{equation}
 \label{eq:86}
{\bf f}^{-1}(\bm{k}) = {\bf f}^{\dagger}(\bm{k}), 
\end{equation}
in order that the spin-dependent Wannier functions in
Equation~\gl{eq:4} form a complete orthonormal basis in the
superconducting band. The Bloch spinors $\varphi_{\bm{k},m}(\bm{r},t)$
are usual Bloch functions with anti-parallel spins possessing,
however, a $\bm k$ dependent symmetry axis $z_{\bm k}$ defined by the
matrix ${\bf f}(\bm k)$.

Since still we consider a superconducting band, the matrices
$f_{ms}(\bm{k})$ can be chosen in such a way that the spin-dependent
Wannier functions
\begin{equation}
 \label{eq:4}
 w_{m}(\bm{r} - \bm{R}, t) = 
\frac{1}{\sqrt{N}}\sum^{BZ}_{\bm{k}}e^{-i\bm{k} \bm{R}}\varphi_{\bm{k},m}(\bm{r},t)
\end{equation}
are optimally localized and symmetry-adapted to the full space group
of the considered metal~\cite{theoriewf}. The sum in
Equation~\gl{eq:4} is over the $N$ vectors $\bm{k}$ of the first
Brillouin zone (BZ), and $\bm{R}$ denotes a lattice vector.  However,
the matrices $f_{ms}(\bm{k})$ cannot be chosen independent of ${\bm
  k}$ since as mentioned in Section~\ref{sec:introduction}, we cannot
unitarily transform the Bloch functions of the superconducting band
into usual (i.e., spin-independent) Wannier functions that are also
optimally localized and symmetry-adapted. Hence, the spin-dependent
Wannier functions differ substantially from usual spin-independent
Wannier functions even if we neglect spin-orbit effects.

The Bloch spinors
may be calculated from the spin-dependent Wannier functions by the
equation
\begin{equation}
  \label{eq:5}
 \varphi_{\bm{k},m}(\bm{r},t) = 
\frac{1}{\sqrt{N}}\sum^{BvK}_{\bm{R}}e^{i\bm{k} \bm{R}}w_{m}(\bm{r} - \bm{R}, t),
\end{equation}
where the sum now is over the $N$ lattice vectors $\bm{R}$ of the
Born-von K\`arm\`an volume (BvK).
\subsection{Nonadiabatic interaction producing {\bf\em k}-space magnetism}
\label{sec:interactionksm}
Let be the operator
\begin{equation}
  \label{eq:6}
  H = H_{HF} + H_{Cb}
\end{equation}
the Hamiltonian in the superconducting band with $H_{HF}$ and
\begin{eqnarray}
H_{Cb} &=& \sum_{\bm R, m}\langle\bm R_{1}, m_1; \bm R_{2}, m_2|H_{Cb}|
\bm R_{1}', m_1'; \bm R_{2}', m_2'\rangle\nonumber\\
&&\times c_{\bm R_{1}m_{1}}^{\dagger}
c_{\bm R_{2}m_{2}}^{\dagger}
c_{\bm R_{2}'m_{2}'}
c_{\bm R_{1}'m_{1}'}
\label{eq:11}
\end{eqnarray}
representing the Hartree-Fock and Coulomb energy, respectively.  The fermion
operators $c_{\bm Rm}^{\dagger}$ and $c_{\bm Rm}$ create and annihilate
electrons in the localized states $|\bm R, m\rangle$
represented by the spin-dependent Wannier functions $w_m(\bm r - \bm
R, t)$ in Equation~\gl{eq:4}.
We write $H_{Cb}$ as
\begin{equation}
H_{Cb} = H_{c} + H_{ex} + H_{z},
\label{eq:12}
\end{equation}
where $H_{c}$ and $H_{ex}$ contain the matrix
elements of $H_{Cb}$ with
\begin{equation}
\bm R_{1} = \bm R_{1}' ~~\mbox{and}~~ \bm R_{2} = \bm R_{2}',
\label{eq:13}
\end{equation}
and
\begin{equation}
\bm R_{1} = \bm R_{2}'~~ \mbox{and}~~ \bm R_{2} = \bm R_{1}',
\label{eq:14}
\end{equation}
respectively, and $H_{z}$ comprises the remaining (non-diagonal) matrix
elements  with
\begin{equation}
\{\bm R_1,\bm R_2\} \neq \{\bm R_1',\bm R_2'\}.
\label{eq:15}
\end{equation}
The operators $H_c$ and $H_{ex}$ represent the Coulomb repulsion and
the exchange interaction, respectively, between atomic-like electrons
and, hence, do not contradict the picture of localized electron states
moving as Bloch waves through the crystal.  $H_z$, on the other hand,
represents an interaction {\em destroying} the atomic-like
motion~\cite{enhm}.

Now consider the operator
\begin{equation}
  \label{eq:16}
  H' = H_{HF} + H_c + H_{ex} = H - H_z
\end{equation}
being gained from $H$ in Equation~\gl{eq:6} by putting $H_z$ equal to
zero, and assume the {\em exact} ground states $|G\rangle$ and
$|G\,'\rangle$ of $H$ and $H'$, respectively, to be determined. The
first postulate of the NHM states that a pure atomic-like motion is
energetic more favorable than an atomic-like motion disturbed by
$H_z$,
\begin{equation}
  \label{eq:17}
  \langle G|H|G\rangle > \langle G\,'|H'|G\,'\rangle,
\end{equation}
if the superconducting band is one of the narrowest bands in the
considered metal, see the detailed substantiation in Ref.~\cite{enhm}.

The second postulate of the NHM states that the electronic transitions
represented by $H_z$ are attributed to the adiabatic approximation and
do not occur in the true nonadiabatic system,
\begin{equation}
  \label{eq:18}
  \langle\bm R_{1}, m_{1}, n; 
\bm R_{2}, m_{2}, n|H_{Cb}|
\bm R_{1}', m_{1}', n; \bm R_{2}', m_{2}', n \rangle
= 0,
\end{equation}
for
\begin{equation}
  \label{eq:19}
  \{\bm R_1,\bm R_2\} \neq \{\bm R_1',\bm R_2'\}
\end{equation}
if Inequality~\gl{eq:17} is true. At the transition to the
nonadiabatic system, the electron system lowers its total Coulomb
energy by the ``nonadiabatic condensation energy''
\begin{equation}
  \label{eq:22}
  \Delta E =  \langle G|H|G\rangle - \langle G\,'|H'|G\,'\rangle.
\end{equation}

Equation~\gl{eq:18} is suggested by the fact that the non-diagonal
matrix elements of $H_{Cb}$ depend very sensitive on the exact form of
the localizes orbitals and, hence, only small modifications should be
required to suppress the transitions represented by $H_z$. The
modified localized orbitals cannot be described within the adiabatic
approximation (since here Inequality~\gl{eq:17} is true) but require
the introduction of nonadiabatic localized states
\begin{equation}
  \label{eq:20}
|\bm R, m, \nu \rangle,  
\end{equation}
possessing the same symmetry as the spin-dependent Wannier functions,
see the detailed discussion in Ref.~\cite{enhm}. The new quantum
number $\nu$ labels the nonadiabatic motion of the atomic
core following the motion of the localized electron, and $\nu = n$
labels the special states satisfying Equation~\gl{eq:18}.

The nonadiabatic symmetry operators (as defined in Equation (B9) of
Ref.~\cite{enhm}) no longer act on the electronic coordinates alone,
but additionally on the coordinate describing that part of the motion
of the atomic core that follows the motion of the electron. Thus, the
electronic motion in the nonadiabatic localized states $|\bm R, m, \nu
\rangle$ is not so confined by symmetry as in the adiabatic states
$|\bm R, m\rangle$. The electrons now move in a potential depending on
which of the adjacent localized states are occupied and on the present
positions of these electrons. Hence, the nonadiabatic localized states
represent a {\em strongly correlated} atomic-like motion.

It is essential that the NHM does not only neglect $H_z$ but
postulates a nonadiabatic mechanism suppressing the transitions
generated by $H_z$. This has the important consequences that, first,
the nonadiabatic Hamiltonian commutes with the operators of the space
group if and only if the nonadiabatic localized states are adapted to
the symmetry of the space group~\cite{enhm}, and, second, the naked
electrons no longer have exact Fermi character. Now, the Fermi
excitations are represented by electrons occupying the nonadiabatic
states $|\bm R, m, n \rangle$ traveling as Bloch states through the
crystal.

The nonadiabatic states are postulated to interpret
Inequality~\gl{eq:17} and to understand Equation~\gl{eq:18}. I believe
that it would be physically needless to try to determine explicitly
the highly complex localized functions representing the nonadiabatic
states. We may assume that the modifications of the adiabatic
electronic orbitals required in Equation~\gl{eq:18} are so small that
any calculation of expectation values (i.e., of diagonal matrix
elements) still can be performed within the adiabatic
approximation. That means that any expectation value in the
superconducting band can be determined in close approximation by
replacing the nonadiabatic localized functions by the adiabatic
spin-dependent Wannier functions~\cite{enhm}.

This has the consequences that, first, the spin-dependent Wannier
functions must be adapted to the symmetry of the space group in order
that the nonadiabatic Hamiltonian correctly commutes with the
operators of the space group, and, second, the expectation
values of the electronic spin directions are determined by the Bloch
spinors in Equation~\gl{eq:5} because they represent the nonadiabatic
Bloch states within the adiabatic approximation. Thus, the adiabatic
Bloch spinors~\gl{eq:5} define the spin direction of the electrons in
the nonadiabatic system, and, consequently, produce $\bm k$-space
magnetism.

In summary, the electrons in a narrow, roughly half-filled
superconducting band may lower their Coulomb energy by the
nonadiabatic condensation energy $\Delta E$~\gl{eq:22} by producing
$\bm k$-space magnetism as described in Section~\ref{sec:ksm}. The
$\bm k$ dependent spin directions are defined by the matrices
$f_{ms}(\bm{k})$ in Equation~\gl{eq:1} which in turn are determined by
the demand that the spin-dependent Wannier functions must be optimally
localized and symmetry-adapted to the space group of the considered
metal.

\section{Discussion}
\label{sec:discussion}
The aim of this paper was to give a graphic description of the
nonadiabatic mechanism of Cooper pair formation defined within the
NHM. The presented picture clearly shows the peculiar features of the
Cooper pair formation within a superconducting band: first, the
postulates of the NHM suggest that the strongly correlated atomic-like
motion in a narrow, roughly half-filled superconducting band produces
$\bm k$-space magnetism in the nonadiabatic system (as described in
Section~\ref{sec:ksm}), and, secondly, at sufficiently low
temperatures the $\bm k$-space magnetism produces Cooper pairs in
turn. This picture clearly demonstrates that the formation of Cooper
pairs produced by $\bm k$-space magnetism shows a great resemblance,
but also a striking difference as compared with the familiar BCS
mechanism~\cite{bcs}. On the one hand, the formation of Cooper pairs
is still mediated by bosons but, on the other hand, the electrons {\em
  necessarily} form Cooper pairs below a transition temperature.  This
necessity of the Cooper pair formation we compare with the effect of
constraining forces in classical systems and, consequently, we speak
of constraining forces stabilizing the Cooper pairs~\cite{josn} or,
more illustratively, of ``spring mounted Cooper
pairs''~\cite{josi}. There is evidence that these constraining forces
are essential for the formation of Cooper pairs, see, e.g., the
Introduction of Ref.~\cite{bi}. In this context, the question whether
or not there exists an attractive interaction between the electrons is
of secondary importance.

\acknowledgments{I am very indebted to Guido Schmitz for his support
  of my work.}



\end{document}

%% file: Figur.tex
\begin{figure*}
\centering
\begin{minipage}[b]{.2\textwidth}
\centering
\includegraphics[width=.65\textwidth,angle=0]{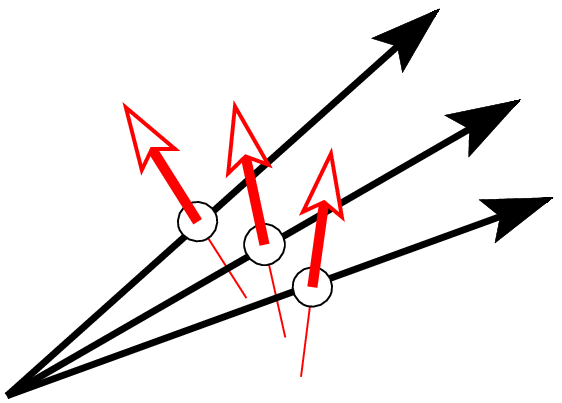}%
\begin{center}
{\bf (a)}
\end{center}
\end{minipage}
\begin{minipage}{.45\textwidth}
\centering
\includegraphics[width=.9\textwidth,angle=0]{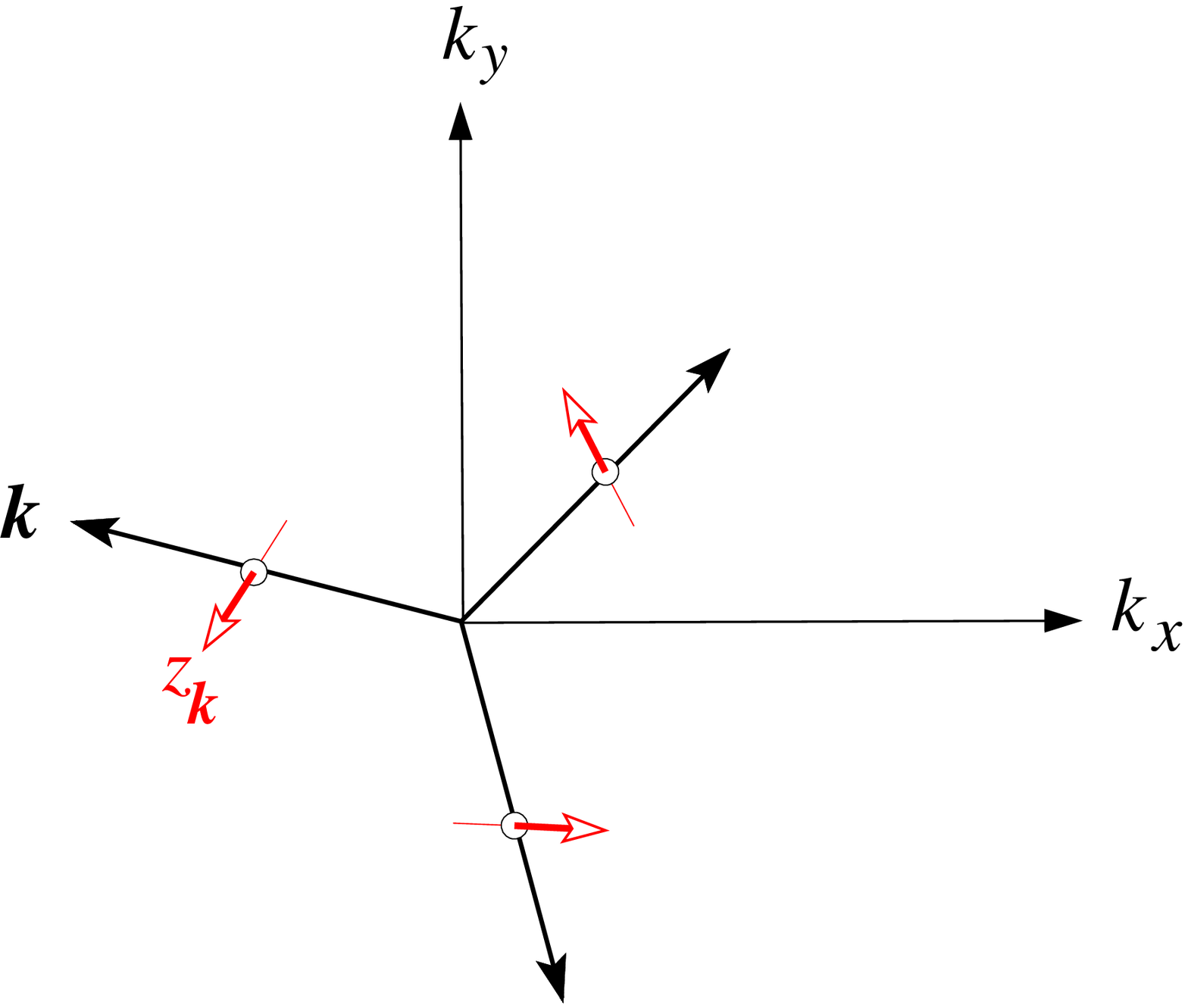}%
\begin{center}
{\bf (b)}
\end{center}
\end{minipage}
\caption{Visualization of the ${\bm k}$ space magnetism in a narrow,
  roughly half-filled superconducting band: the black arrows show the
  ${\bm k}$ vectors of Bloch electrons moving in general positions at
  the Fermi level, and the red arrows indicate the symmetry axis
  $z_{\bm k}$ of the spin of the Bloch electron with wave vector $\bm
  k$. The $z_{\bm k}$ axes generally intersect the drawing plane.  Figure
  {\bf (a)} demonstrates that the $z_{\bm k}$ axes change continuously
  in $\bm k$ space. Figure {\bf (b)} shows the $\bm k$ vectors of three
  Bloch electrons connected by symmetry (in a crystal with the
  hexagonal space group P3) and demonstrates that also the $z_{\bm k}$
  axes are connected by symmetry.}
\label{fig:zeichnung}
\end{figure*}
